\documentclass[%
 preprint,
superscriptaddress,
 amsmath,amssymb,
 aps,
prstab,
]{revtex4-1}

\usepackage{graphicx}
\graphicspath{{figure/}}
\usepackage{overpic}
\usepackage{dcolumn}
\usepackage{bm}
\usepackage{float}
\usepackage{algorithm}
\usepackage{algorithmic}
\usepackage{xcolor}
\usepackage{setspace}

\begin{document}

\title{Neural-network-based longitudinal electric field prediction in nonlinear plasma wakefield accelerators}

\author{Xiaoning Wang}
\affiliation{Institute of High Energy Physics,
Chinese Academy of Sciences, Beijing 100049, China}
\affiliation{Spallation Neutron Source Science Center, Dongguan 523803, China}

\author{Ming Zeng}
\email[Corresponding author: ]{zengming@ihep.ac.cn}
\affiliation{Institute of High Energy Physics,
Chinese Academy of Sciences, Beijing 100049, China}
\affiliation{University of Chinese Academy of Sciences, Beijing 100049, China}
\affiliation{Center for High Energy Physics, Henan Academy of Sciences, Zhengzhou 450046, China}

\author{Dazhang Li}
\affiliation{Institute of High Energy Physics,
Chinese Academy of Sciences, Beijing 100049, China}
\affiliation{Center for High Energy Physics, Henan Academy of Sciences, Zhengzhou 450046, China}

\author{Weiming An}
\email[Corresponding author: ]{anweiming@bnu.edu.cn}
\affiliation{School of Physics and Astronomy, Beijing Normal University, Beijing 100875, China}
\affiliation{Institute for Frontiers in Astronomy and Astrophysics, Beijing Normal University, Beijing 102206, China}

\author{Wei Lu}
\affiliation{Institute of High Energy Physics,
Chinese Academy of Sciences, Beijing 100049, China}
\affiliation{Center for High Energy Physics, Henan Academy of Sciences, Zhengzhou 450046, China}

\begin{spacing}{1.2}
\begin{abstract}
Plasma wakefield acceleration holds remarkable promise for future advanced accelerators. The design and optimization of plasma-based accelerators typically require particle-in-cell simulations, which can be computationally intensive and time consuming. In this study, we train a neural network model to obtain the on-axis longitudinal electric field distribution directly without conducting particle-in-cell simulations for designing a two-bunch plasma wakefield acceleration stage. By combining the neural network model with an advanced algorithm for achieving the minimal energy spread, the optimal normalized charge per unit length of a trailing beam leading to the optimal beam-loading can be quickly identified. This approach can reduce computation time from around 7.6 minutes in the case of using particle-in-cell simulations to under 0.1 seconds. Moreover, the longitudinal electric field distribution under the optimal beam-loading can be visually observed. Utilizing this model with the beam current profile also enables the direct extraction of design parameters under the optimal beam-loading, including the maximum decelerating electric field within the drive beam, the average accelerating electric field within the trailing beam and the transformer ratio. This model has the potential to significantly improve the efficiency of designing and optimizing the beam-driven plasma wakefield accelerators.
\end{abstract}
\end{spacing}


\maketitle

\section{Introduction}
In the current domain of accelerator science, design and optimization mainly rely on the combination of theoretical analysis and numerical simulations. During the operational phase of an accelerator, hardware adjustments often depend on feedback from the beam position monitor (BPM) and associated instruments. In recent years, machine learning, which is rapidly developing, has already been applied to accelerator design and operational control, resulting in remarkable advancements~\cite{RN163,RN162,RN166,RN169,RN170,RN306,2023ACCML,2024ACCML}. The application of machine learning algorithms has enabled improvements in design choices and operational strategies. A model-independent extremum seeking method is introduced to control the longitudinal phase space of an electron beam in the compression section. This method has been successfully implemented in the Linac Coherent Light Source, achieving control resolutions on the femtosecond scale~\cite{RN163}. \textcite{RN166} have proposed a feedforward correction method based on a neural network to reduce beam size fluctuations to the experimental noise level of $\rm0.2\ \mu m$, demonstrating higher efficiency compared with traditional feedback and feedforward methods. A machine learning-based approach has been proposed to predict particle accelerator failures using beam current measurements, achieving up to 92\% accuracy in classifying bad pulses and identifying failures before they occur at the Oakridge Spallation Neutron Source~\cite{RN169}. In addition, this application of machine learning techniques has promoted more detailed methods to optimize accelerator performance, leading to enhancements in efficiency and reliability. Scientists have developed an online multi-time scale and multi-objective optimization algorithm that enables real-time feedback control in particle accelerators, effectively reducing beam emittance and stabilizing beam trajectories in Advanced Proton Driven Plasma Wakefield Acceleration Experiment~\cite{RN170}. A neural network model is employed to predict electron beam trajectories, achieving a prediction accuracy that reaches the BPM resolution limit under high charge conditions~\cite{RN306}. 

Novel acceleration methods based on laser or beam-driven plasma wakefields have demonstrated remarkable outcomes in terms of attainable gradients, easily exceeding the GeV/m threshold in plasma~\cite{RN88,RN8}. This development has led to widespread interest in plasma wakefield acceleration within the accelerator community~\cite{RN122,RN121,RN120,RN113,RN111,RN114,RN115,RN80,RN82,RN131,RN152,RN11,RN13,RN117,RN318,RNLPS,RN262,RNEp1,RN255,RN257,RN259,2023LWFA,2024LWFA}. In recent years, the application of machine learning in the field of plasma wakefield acceleration has attracted substantial attention and has achieved a series of outcomes~\cite{2020BESLWFA,RN209,2021SMLLWFA,2021BESLWFA,2023MLLWFABE,2023PNN,2023BEL,2024BELWFA,2024NNLWFA}. This approach has been increasingly integrated into the analysis and optimization of plasma wakefield accelerators, improving our ability to unravel the complexities involved and to develop advanced accelerator technologies. These advancements highlight the transformative potential of machine learning in enhancing the efficiency and effectiveness of the plasma wakefield acceleration research.

Given the complex nature of the physical phenomena, the design of a plasma wakefield accelerator requires the employment of high-fidelity particle-in-cell (PIC) simulations. PIC simulations calculate the self-consistent interactions between particles and electromagnetic fields with minimal assumptions. However, the considerable computational demands of these simulations make it impractical to conduct an optimization over a vast array of parameters. Exploiting more efficient methods to optimize the design of plasma wakefield accelerators is crucial to unlocking their full potential. In addition to the ongoing improvement of computational performance, the number of simulations required for optimization can be reduced using intelligent algorithms. Advanced algorithms such as Bayesian optimization~\cite{BES,2020BESLWFA,2021BESLWFA,2023MLLWFABE,2023BEL,2024BELWFA}, neural network models~\cite{loss_intro,RN231,RN209,2024NNLWFA} and the Broyden-Fletcher-Goldfarb-Shanno (BFGS) algorithm ~\cite{1967Quasi,RN195} have effectively improved the design and optimization of plasma wakefield accelerators. 

In the bubble regime~\cite{RN80}, the longitudinal electric field of the wake is altered upon the loading of the trailing beam. Proper loading can flatten the longitudinal electric field within the trailing beam, enabling uniform acceleration of beam particles and minimizing the energy spread (i.e.\ optimal beam-loading~\cite{RN152}), which is crucial to accelerator applications. The effect of beam-loading significantly influences the beam quality and has been a subject of extensive research~\cite{RNN76,RN75,RN152,RNEp1,RNEp2,RN209,RN195,2023RNE,2023MLLWFABE}. Although intelligent algorithms can enhance the optimization efficiency, this process typically rely on PIC simulations to obtain the longitudinal electric field. In this study, we present a novel method of neural networks to directly obtain the on-axis longitudinal electric field under the beam-loading effect. Moreover, a method based on the BFGS algorithm, as introduced in a previous study~\cite{RN195}, effectively minimizes the energy spread of a trailing beam, which is crucial for the design of a two-bunch beam-driven plasma wakefield acceleration (PWFA). Two fitting formulas of beam's normalized charge per unit length $\Lambda=n_b\sigma_r^2$, where $n_b$ is the beam peak density and $\sigma_r$ is the rms beam spot size, and the transformer ratio $R$ under the optimal beam-loading, which enhance the efficiency of the optimization process, are provided by \textcite{RN195}. In the present study, we propose an alternative novel approach in which the utilization of our machine learning model with the BFGS algorithm can accelerate the achievement of optimal beam-loading. In addition, more specific information of design parameters can be obtained directly through our model. In Sec.~\ref{NN}, a neural network model for predicting the on-axis longitudinal electric field in the bubble regime is presented. Afterwards, we use this model to efficiently achieve the optimal beam-loading, while also acquiring field information and transformer ratio in Sec.~\ref{app_NN}. Finally, we conclude and analyze the results presented in this study. Throughout this paper, we adopt the normalized units to eliminate the dependence on plasma density, where the beam density is normalized to the plasma density $n_p$, velocities to the spreed of light $c$, charge to the electron charge $e$, mass to the electron mass $m_e$, length to $k_p^{-1} \equiv c/\omega_p$ and electric field to $m_ec\omega_p/e$, where $\omega_p = \sqrt{4\pi e^2n_p/m_e}$ is the plasma frequency. In fact, parameters in regular units can be easily obtained if a specific plasma density is used.

\section{Prediction of Longitudinal Electric Field Distribution with Neural Networks\label{NN}}
\subsection{Construction and Training of Neural Network Models\label{NN_process}}
\newcolumntype{C}[1]{>{\centering\arraybackslash}p{#1}}
Neural networks, composed of simple nodes with adaptability, can fit various nonlinear relationships. In our work, we utilize a general fully connected feedforward neural network model to predict the on-axis longitudinal electric field distribution, because this issue does not involve high-dimensional reduction \cite{RN232} or temporal problems \cite{2001RNN}. 

A simple fully connected feedforward neural network model consists of an input layer, a hidden layer, and an output layer. The neurons in the input layer receive input data, which is processed by the neurons in the hidden and output layers. The output layer is responsible for outputting the final result. Activation functions are typically introduced to create a nonlinear mapping between the input and the output, enhancing the network's generalization capability to fit various complex relationships~\cite{activation}. During the model training, model parameters are usually updated using the back-propagation method \cite{back-propagation}. Back-propagation calculates the gradient of the empirical risk (average loss of the training dataset) with respect to the parameters of each layer's activation function and adjusts the parameters using gradient descent to improve the fitting performance. At the same time, it is necessary to select appropriate optimization algorithms~\cite{opti_al,SGD_intro,Adam_ref} to solve problems iteratively, which can improve the prediction accuracy.

\begin{figure}
    \centering
    \begin{overpic}
    [width=0.9\textwidth]{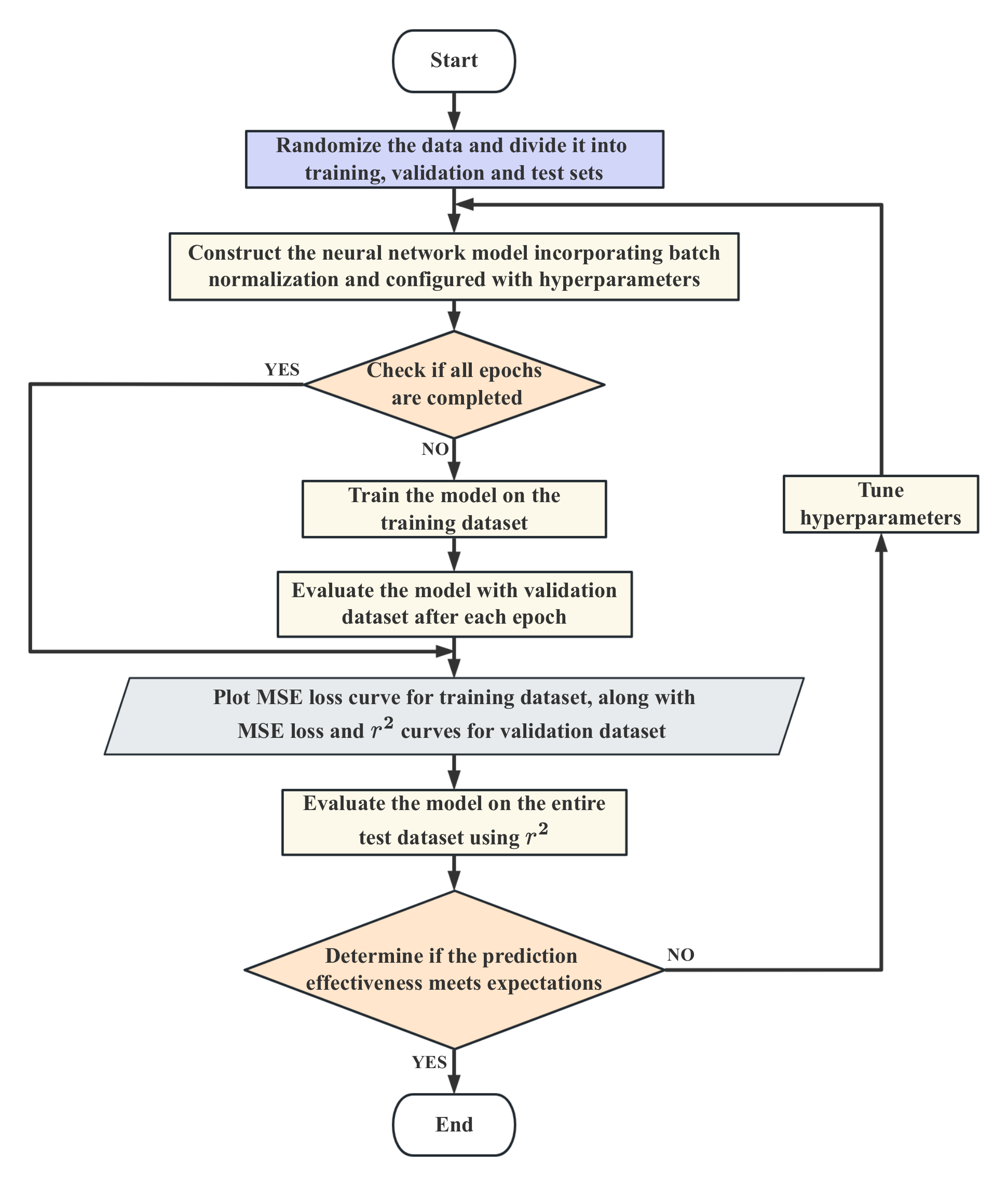}
    \end{overpic}
\renewcommand{\baselinestretch}{1.5}
 \caption{Flowchart of the Neural Network model construction and training process.}\label{NN_flowchat}
\end{figure}

The general process for constructing and training a neural network model is illustrated in Fig.~\ref{NN_flowchat}. Before model training, it is common to randomize the data and divide it into training, validation and test datasets with roughly equal proportions. Then, the neural network model incorporating batch normalization method \cite{BN} is constructed with hyperparameters~\cite{hyper-par}, which will be continuously modified to achieve satisfactory predictive performance as depicted in Fig.~\ref{NN_flowchat}. Subsequently, the model is trained on the training dataset, with the number of iterations determined by the specified number of epochs, and the mean squared error (MSE) loss function~\cite{loss_intro} value is recorded. After each epoch, the model is evaluated on the validation dataset (without batch division) to prevent overfitting~\cite{RN200}, calculating the MSE loss and the coefficient of determination $r^2$ \cite{RN200} for each epoch. The metric $r^2$ measures how well unseen samples tend to be predicted by the model. It is defined as $r^2(y,y_p) = 1-\frac{\sum_{i=1}^N(y_i-y_{pi})^2}{\sum_{i=1}^N(y_i-\bar{y})^2} $, where $\bar{y} = \frac{1}{N}\sum_{i=1}^N y_i$, $y_i$ is the corresponding sample's value for total $N$ unseen samples and $y_{pi}$ is the predicted value of the $i-$th sample. The closer $r^2$ is to 1, the better the goodness of fit is. Once the model has completed the specified number of epochs, curves showing the change with epochs for the training and validation datasets are obtained. These curves play an important role in the model training, helping to avoid overfitting and tune hyperparameters. Ideally, the loss curve should show a decrease and smooth stabilization for the training and validation datasets, while the accuracy curve for the validation dataset should exhibit an increase and smooth stabilization. Finally, the model is evaluated using the entire test dataset based on $r^2$.

\subsection{The Longitudinal Electric Field Distribution Prediction Model}
\begin{figure}
    \centering
    \begin{overpic}
    [width=0.65\textwidth]{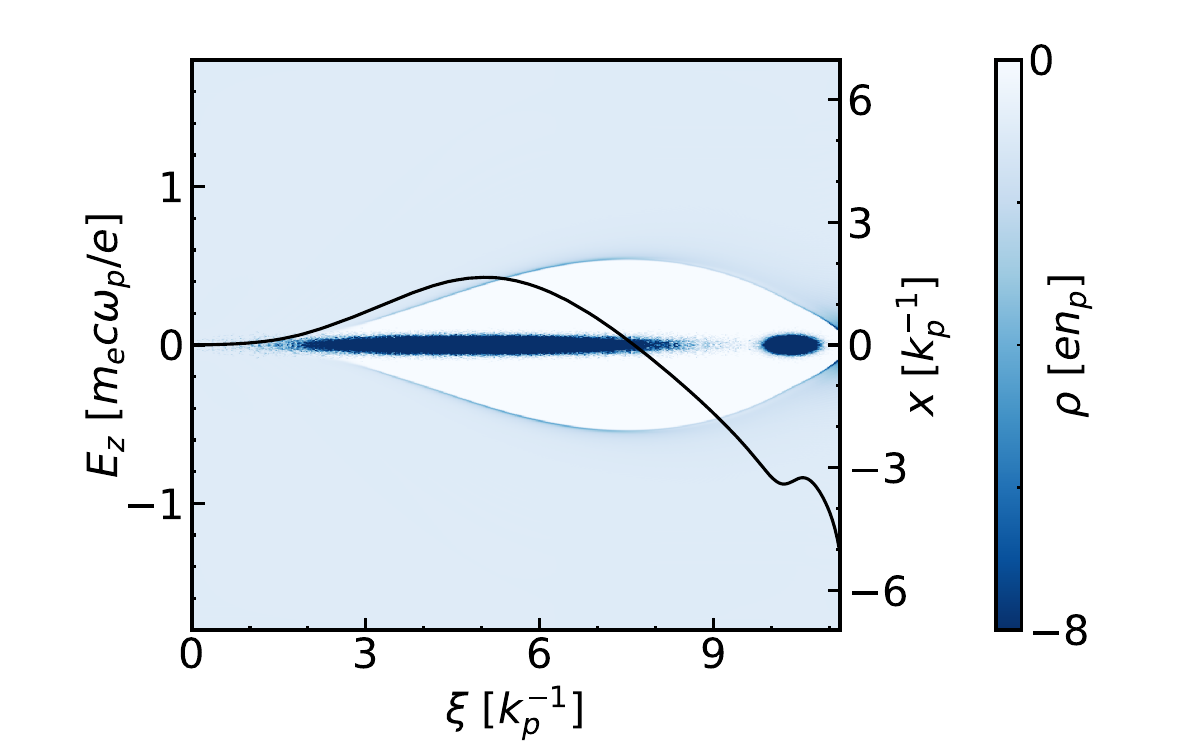}
    \end{overpic}
\renewcommand{\baselinestretch}{1.5}
 \caption{Schematic of a beam-driven plasma wakefield acceleration. The beam in the front is the drive beam, and the one in the back is the trailing beam. The black line is the on-axis longitudinal wakefield. Parameters in this simulation are $\Lambda_d = 0.90$, $\Lambda_t = 1.26$, $\sigma_{zd} = 1.40$, $\sigma_{zt} = 0.20$ and $d =5.3$. The simulation box has the size of $14.0\times14.0\times11.0$ and contains $1024\times1024\times512$ cells (in $x$, $y$ and $\xi$ directions, respectively).}\label{PBA}
\end{figure}

In the bubble regime of the beam driven plasma wakefield, a charged particle beam acting as the drive beam excites the 3D nonlinear wake as shown in Fig.~\ref{PBA}. The other electron beam serving as the trailing beam is loaded at the appropriate phase inside the wake for acceleration~\cite{RN152}. The tri-Gaussian distribution is a widely accepted profile for the beams, which has the density distribution $\rho_b = n_b\cdot\exp(-\frac{x^2+y^2}{2\sigma_r^2})\exp(-\frac{\xi^2}{2\sigma_z^2})$, where $\xi=ct-z$ is the co-moving coordinate, $x$ and $y$ are the transverse coordinates and $\sigma_z$ is the rms beam length. In a PWFA with tri-Gaussian beams, the drive (trailing) beam's current distribution can be determined by its normalized charge per unit length $\Lambda_d$ ($\Lambda_t$), and rms beam length $\sigma_{zd}$ ($\sigma_{zt}$). By combining these parameters with the beam separation $d$ which is defined as the distance between the center of the drive beam and that of the trailing beam, the longitudinal wakefield distribution under the beam-loading can be determined. Additionally, to define the prediction interval, the input variables also include the center position of the drive beam $C_d$ and the longitudinal simulation box length $L_z$, resulting in a total of 7 input variables. The initial dataset for our model is derived from the data obtained through a large-range parameter scanning. This process is conducted to minimize the energy spread of trailing beams using the BFGS algorithm with PIC simulations~\cite{RN195}. Our dataset contains data from the iterative process of achieving the optimal beam-loading. In this optimization problem, the beam spot size $\sigma_r$ is neglected under the condition $R_b\gg\sigma_r$ where $R_b$ is the bubble radius~\cite{RN210,RN195}. Under this condition, any change in the beam spot size, while maintaining the charge per unit length $\Lambda = n_b\sigma^2_r$ and the beam length constant, will have minimal impact on the wake~\cite{RN210,RN80}. Parameters $\Lambda_d$, $\sigma_{zd}$, $\sigma_{zt}$ and $d$ are fixed to find the optimal $\Lambda_t$.

\begin{figure}
    \centering
    \begin{overpic}
    [width=0.495\textwidth]{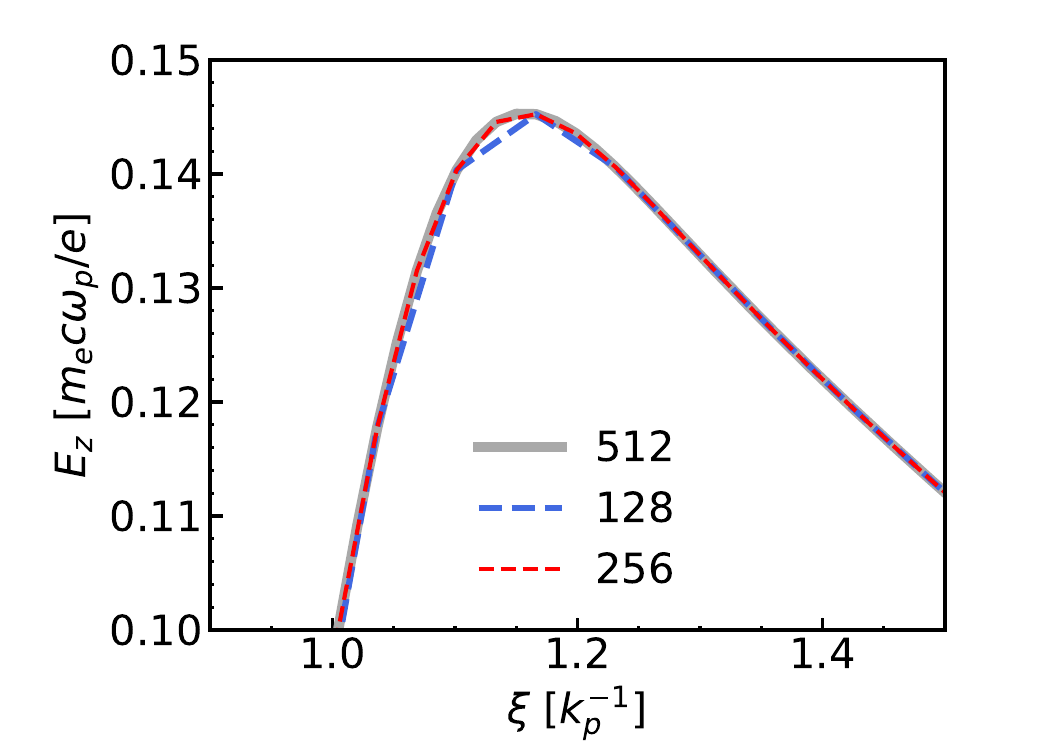}
    \put(21,61){\small{(a)}}
    \end{overpic}
    \begin{overpic}
    [width=0.495\textwidth]{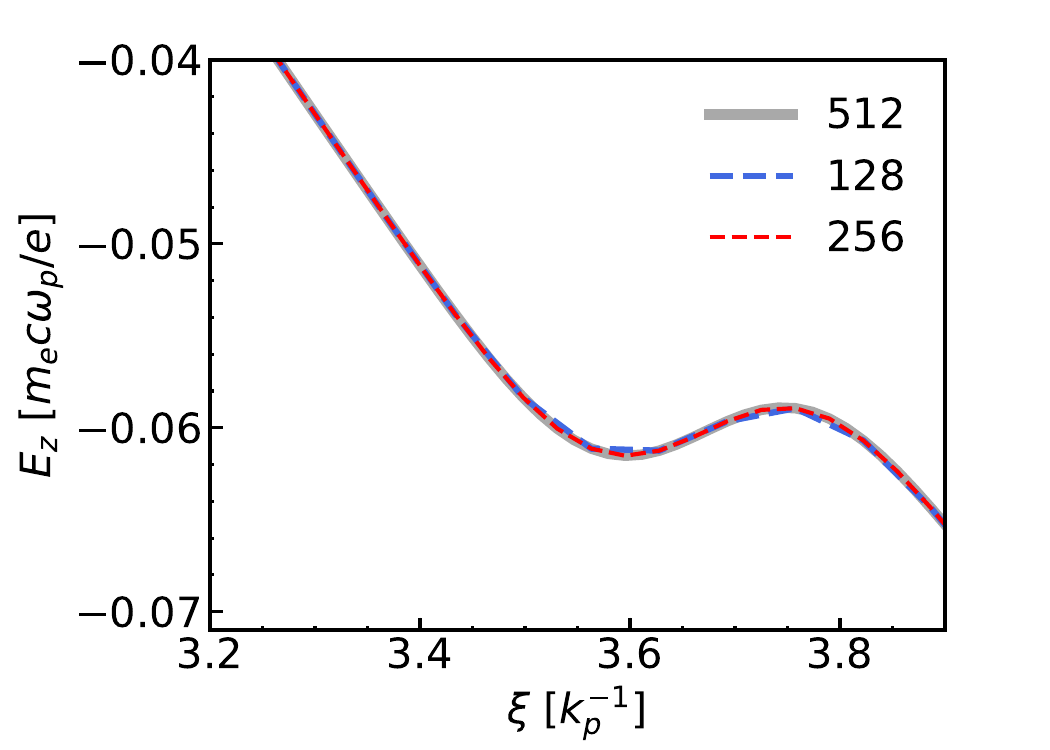}
    \put(21,61){\small{(b)}}
    \end{overpic}
\renewcommand{\baselinestretch}{1.5}
 \caption{Axial distribution of (a) the decelerating wakefield within the drive beam and (b) the accelerating wakefield within the trailing beam. Parameters in this simulation are $\Lambda_d=0.29$, $\sigma_{zd}=0.095$, $\sigma_{zt}=0.095$, $d=2.65$, $\Lambda_t=0.087$, $C_d = 1.0$ and $L_z = 4.13$. The gray, red and blue lines correspond to the output point numbers of 512, 256 and 128, respectively.}\label{Ez_de}
\end{figure}

In order to ensure an adequate number of samples across different ranges, we select a subset of data in the range of $\Lambda_d\leq3.3$ and $\Lambda_t\leq3.3$, which comprises 25335 data sets obtained quickly and reliably from simulations using the quasi-static PIC code QuickPIC~\cite{RN72}. The on-axis longitudinal wakefield contains 512 output points within this range. The parameter ranges, listed in Tab.~\ref{NN_tab}, basically cover the parameters of current PWFA facilities~\cite{RN15,RN14,RN136,RN177} with the plasma density of $\sim 10^{16}\ {\rm cm^{-3}}$. Note that Tab.~\ref{NN_tab} lists the global ranges for $d$, $C_d$ and $L_z$, and the actual ranges of these parameters vary according to the beam parameters. The model prediction difficulty rises with the number of prediction points. To reduce the training difficulty of the model and improve the prediction accuracy, we reduce the number of on-axis $E_z$ output points using a uniform sampling method as shown in Fig.~\ref{Ez_de}. In Fig.~\ref{Ez_de}(a) we show a part of the decelerating wakefield inside the drive beam, while Fig.~\ref{Ez_de}(b) illustrates the accelerating wakefield within the trailing beam. Considering the smoothness of the predictive curve and the ease of the model training, we set the output data points for $E_z$ to be 256 to sample the original 512 grid points.

\begin{table}\small
\caption{\label{NN_tab}
Range of parameters for model training.}
\renewcommand{\arraystretch}{1.3}
{\begin{tabular}{C{0.35\textwidth}C{0.35\textwidth}} \toprule
Parameters & Range \\ \colrule
$\Lambda_d$ & [0.015, 3.30] \\
$\sigma_{zd}\ [k_p^{-1}]$ & [0.0952, 1.90] \\
$d\ [k_p^{-1}]$ & [0.27, 8.58] (the global range)\\
$\sigma_{zt}\ [k_p^{-1}]$ & [0.0952, 1.61]\\
$\Lambda_t$ & [0.011, 3.30] \\
$C_d\ [k_p^{-1}]$ & [1.0, 6.66] (the global range)\\
$L_z\ [k_p^{-1}]$ & [4, 15] (the global range)\\ \botrule
\end{tabular}}
\end{table}

\begin{figure}
    \centering
    \begin{overpic}
        [width=0.65\textwidth]{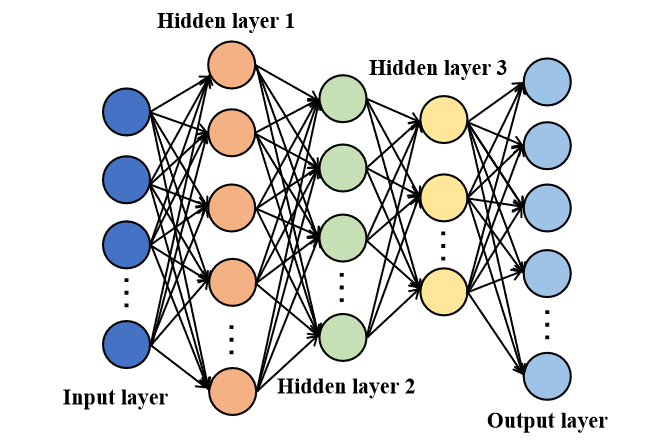}
    \end{overpic}
\renewcommand{\baselinestretch}{1.5}  
\caption{Schematic of the neural network model for the longitudinal wakefield prediction. The input layer consists of 7 neurons and the output layer consists of 256 neurons. This network has three hidden layers with 300, 150 and 50 neurons, respectively.}\label{NN_model}
\end{figure}

\begin{figure}
    \centering
    \begin{overpic}
    [width=0.68\textwidth]{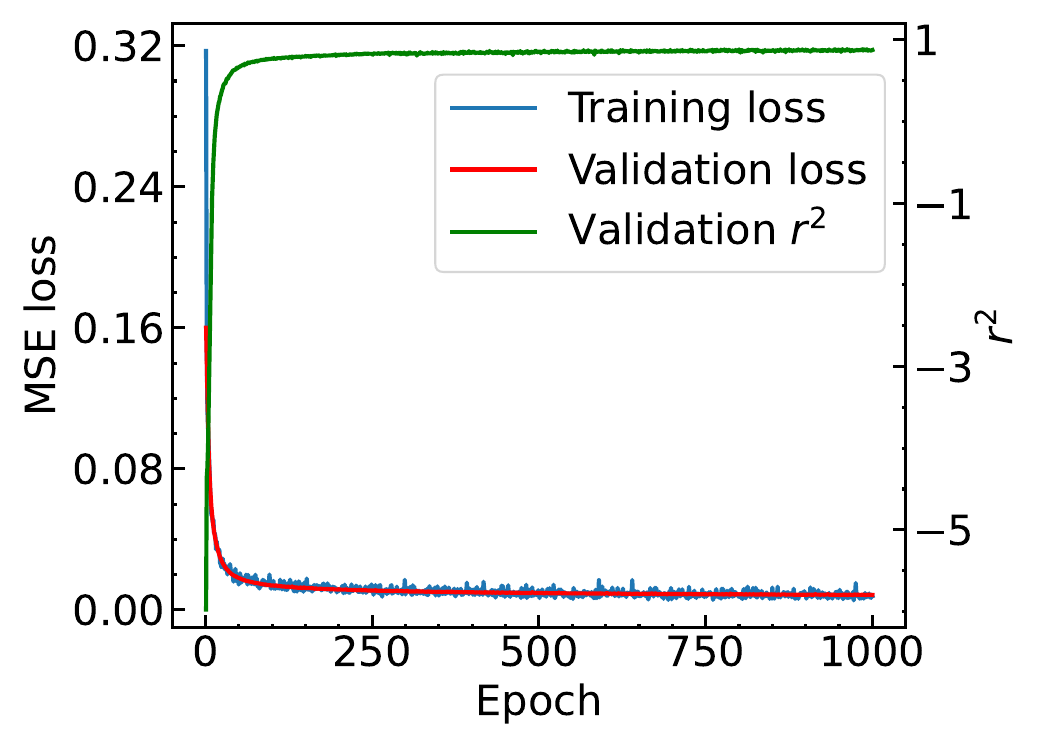}
    \end{overpic}
    \caption{The curves of MSE loss for the training dataset (blue line) and the validation dataset (red line) over epochs, as well as the curve of $r^2$ for the validation dataset (green line) over epochs.}\label{trained_curve}
\end{figure}

Following the model training process introduced in Sec.~\ref{NN_process}, we proceed to build and train a neural network model using the machine learning framework PyTorch \cite{pytorch}. Prior to commencing the model training, we divide the 25335 sets of data falling within the parameter ranges listed in Tab.~\ref{NN_tab} into the training, validation, and test datasets, with proportions of $60\%$, $20\%$, and $20\%$, respectively. The model structure utilized for predicting the on-axis longitudinal wakefield distribution is illustrated in Fig.~\ref{NN_model}. The model consists of one input layer, three hidden layers and one output layer. The number of neurons in the input and output layers is determined by the input variables and output variables. This neural network model has 7 neurons in the input layer and 256 neurons in the output layer. The numbers of neurons in the three hidden layers are 300, 150 and 50, respectively, following a decreasing pattern. The initial layers can learn more low-level features, which are then fed into the subsequent layers to learn relatively higher-level features~\cite{loss_intro}.  The model is first trained using the training dataset and then validated using the validation dataset to prevent overfitting. During the model training process, the ReLU function~\cite{relu} is used as the activation function to enhance the neural network's nonlinear fitting capability. The selected optimization algorithm is the Adam algorithm~\cite{Adam_ref} with a learning rate of LR = 0.0001, and the MSE is selected as the loss function. After 1000 epochs of iteration, the model shown in Fig.~\ref{NN_model} demonstrates good predictive performance. The GPU acceleration is utilized to accelerate the model training, and the total training time on a single GPU (NVIDIA Tesla V100 32G) is around 50 minutes. As illustrated in Fig.~\ref{trained_curve}, the loss curves for the training (blue line) and validation (red line) datasets both show a declining trend, eventually stabilizing over epochs. Meanwhile, the $r^2$ curve (green line) for the validation dataset shows an increase and eventually stabilizes over epochs. The curves of MSE loss and $r^2$ for the trained model demonstrate the expected trends. Finally, the model is evaluated using the test dataset based on the coefficient of determination $r^2$, resulting in $r^2 = 0.90$. These results indicate a good predictive performance of the model.

\begin{figure}
    \centering
    \begin{overpic}
    [width=0.495\textwidth]{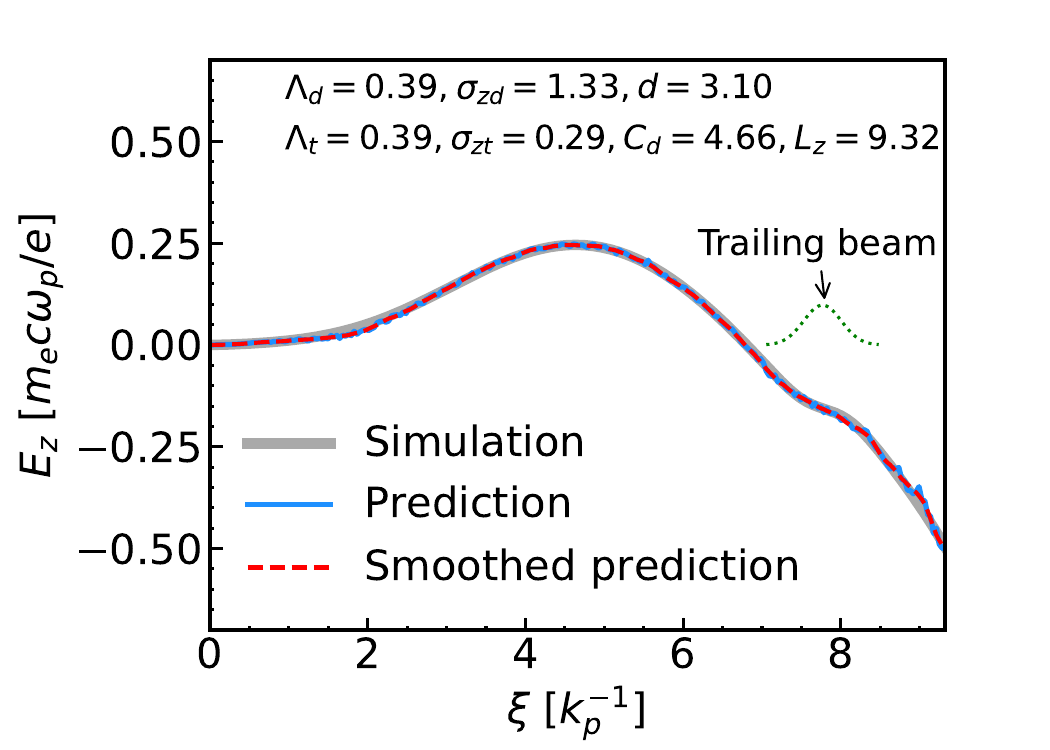}
    \put(21,60.5){\small{(a)}}
    \end{overpic}
    \begin{overpic}
    [width=0.495\textwidth]{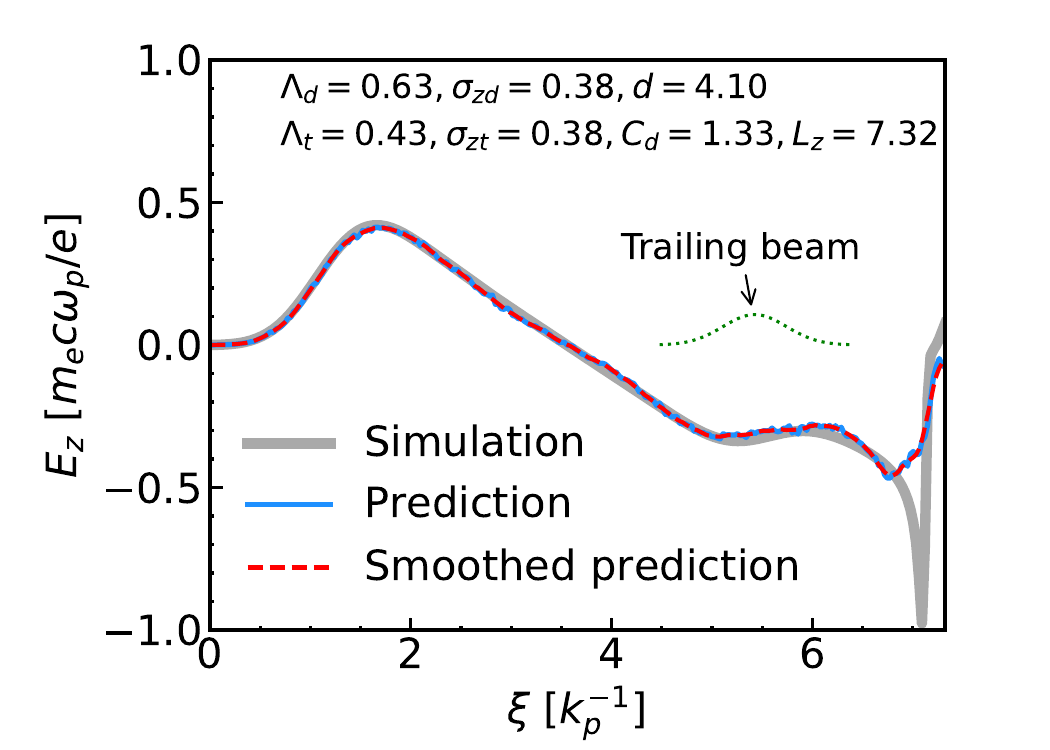}
    \put(20.5,60.5){\small{(b)}}
    \end{overpic}
    \begin{overpic}
    [width=0.495\textwidth]{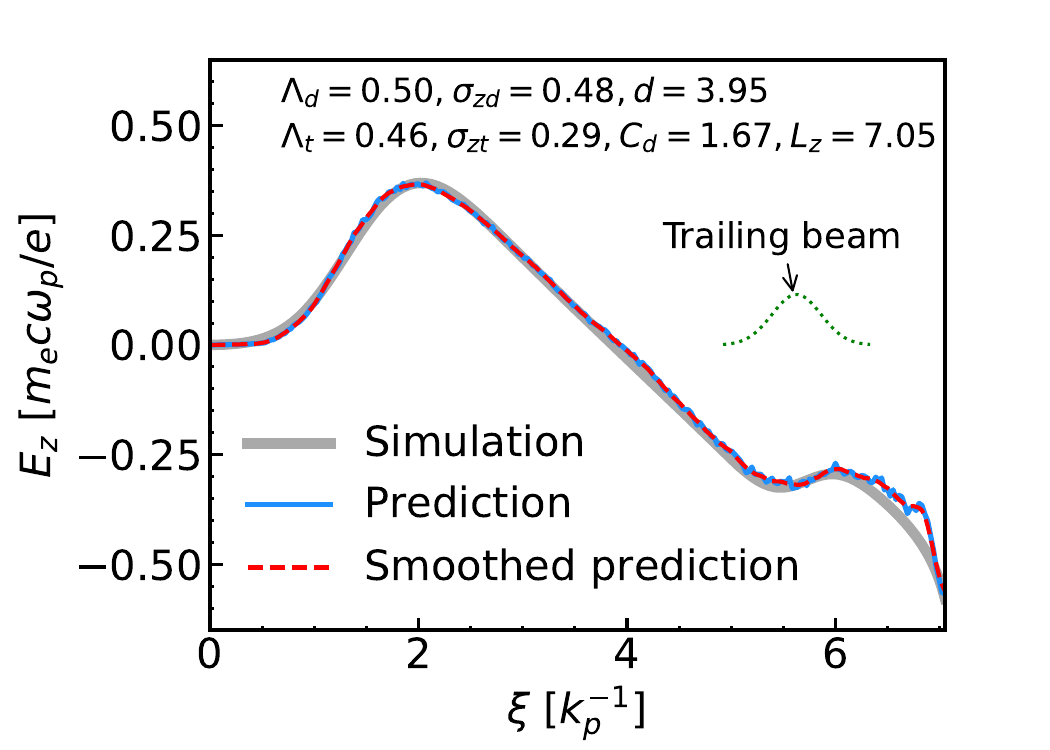}
    \put(21,60.5){\small{(c)}}
    \end{overpic}
    \begin{overpic}
    [width=0.495\textwidth]{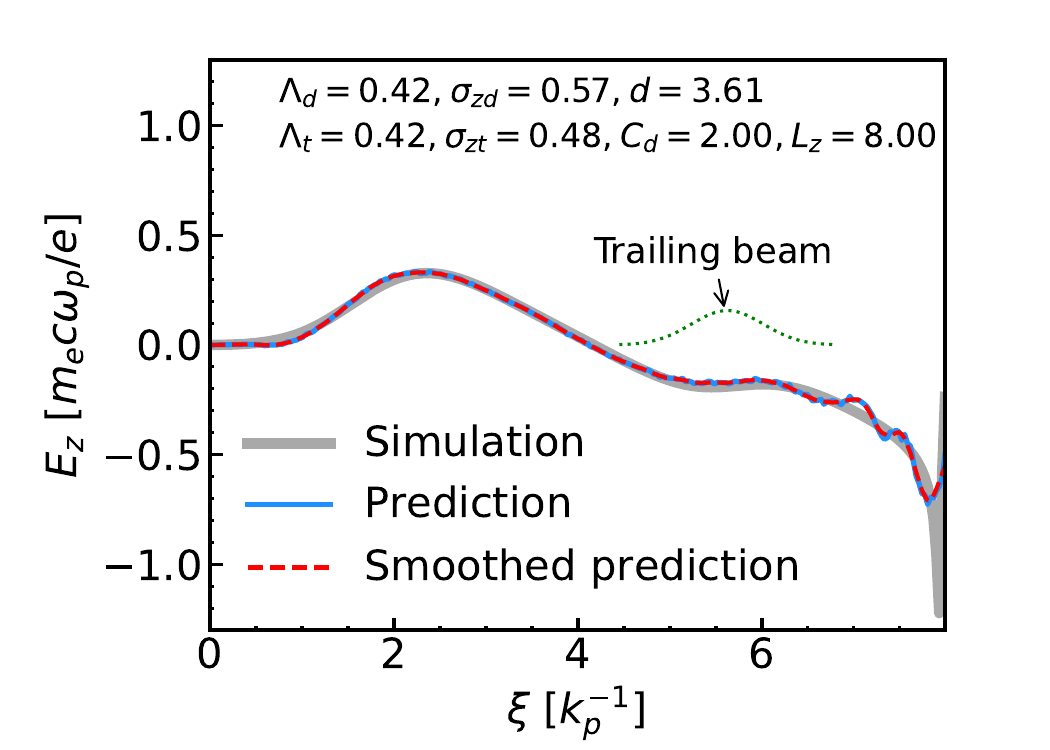}
    \put(20.5,60.5){\small{(d)}}
    \end{overpic}
 \caption{(a)-(d) show the axial distribution of the longitudinal wakefield from the PIC simulation (gray line) and the neural network model (blue line), with the parameters written in each of the subplots. The red dashed line is the smooth of the prediction from the neural network model. The green dashed line is the longitudinal profile of the trailing beam.}\label{Ez_pre}
\end{figure}

In the following, the design with the parameters $\Lambda_d=0.39$, $\sigma_{zd}=1.33$, $\sigma_{zt}=0.29$, $d=3.10$, $\Lambda_t=0.39$, $C_d = 4.66$ and $L_z = 9.32$ is taken as an example to visually demonstrate the predictive performance of the model as shown in Fig.~\ref{Ez_pre} (a). The gray solid line represents the distribution curve obtained from the PIC simulation, and the blue solid line represents the predicted curve of the neural network model, showing a good predictive performance visually. In Fig.~\ref{Ez_pre}(b)-(d), we compare more results from the PIC simulation with those given by our model, which are also in good agreements. To obtain smooth curves of the on-axis $E_z$, we here apply the S-G smoothing method~\cite{RN251} to smooth the predicted curves, which are shown as the red dashed lines in Fig.~\ref{Ez_pre}. By comparing the predicted distribution with the simulated one, we find that there are more noise points in the predicted field behind the trailing beam (green dashed line) compared with other parts. This is not only due to the limited predictive accuracy of the neural network model itself, but also because the field at the tail of the bubble changes more significantly. Nevertheless, the most important information is the $E_z$ field within the ranges of the beams. Therefore, the above prediction deviation has negligible impact on the practical application of the model.

\section{Application of a Neural-network-based Prediction Model}\label{app_NN}
In the bubble regime, when the beam is located inside the axisymmetric bubble, the on-axis longitudinal electric field of the wake is the primary factor affecting the beam energy spread, independent of the beam's transverse position~\cite{RN80}. Therefore, flattening the longitudinal electric field within the trailing beam is the most effective strategy for minimizing the beam energy spread. An efficient optimization method to achieve the optimal beam-loading for minimizing the energy spread of trailing beams has been proposed by \textcite{RN195}. The following function is used as the objective function,
\begin{equation}
    F(\Lambda_t) = \sqrt{\frac{\int_{\xi_s}^{\xi_e}(E_z(\xi))^2\lambda_{bt}(\xi)d\xi}{\int_{\xi_s}^{\xi_e}\lambda_{bt}(\xi)d\xi}-\left(\frac{\int_{\xi_s}^{\xi_e} E_z(\xi)\lambda_{bt}(\xi)d\xi}{\int_{\xi_s}^{\xi_e}\lambda_{bt}(\xi)d\xi}\right)^2},\label{eq1}\nonumber
\end{equation}
where $\xi_s$\ ($\xi_e$) represent the head (tail) location of the trailing beam. And $\lambda_{bt}(\xi) = \int\rho_{bt}(x, y,\xi)dxdy$ denotes the normalized charge per unit length of the trailing beam, where $\rho_{bt}$ is the normalized charge density of the trailing beam and $\Lambda_t$ is the peak value of $\lambda_{bt}(\xi)$. The objective function $F(\Lambda_t)$ is the mean square deviation of weighted on-axis $E_z$, where the density profile of the trailing beam is used as the weight. Given initial parameters $\Lambda_d$, $\sigma_{zd}$, $\sigma_{zt}$ and $d$, the BFGS optimization algorithm can be utilized with the feedback of QuickPIC during the iterations to determine the optimal $\Lambda_t$ for achieving the optimal beam-loading. For the case with $\Lambda_d=1.0$, $\sigma_{zd}=1.0$, $\sigma_{zt}=0.25$ and $d=4.5$, QuickPIC is called 26 times for feedback during the optimization, leading to the optimal solution of $\Lambda_t = 1.49$ in approximately 7 minutes. Then, a fitting formula of $\Lambda_t$ enables the rapid determination of the optimal $\Lambda_t$, providing a quick method for achieving the optimal beam-loading without conducting PIC simulations~\cite{RN195}. As shown in Tab.~\ref{lam_diff}, for the above case, the optimal $\Lambda_t=1.49$ can be calculated directly using the fitting formula of $\Lambda_t$. This method improves the design and optimization efficiency of PWFA, but PIC simulations are still needed to observe the longitudinal wakefield distribution under the optimal beam-loading to verify the effectiveness of the acceleration.

\begin{table}\small
\caption{\label{lam_diff}
Optimal $\Lambda_t$ obtained from the PIC simulation, the fitting formula of $\Lambda_t$ in Ref.~\cite{RN195}, and the neural network model, along with the computational time for each of the three methods.}
\renewcommand{\arraystretch}{1.3}
{\begin{tabular}{C{0.4\textwidth}C{0.2\textwidth}C{0.3\textwidth}} \toprule
        Method & Optimal $\Lambda_t$ & Computational time\\ \colrule
        PIC simulation &  1.49 & $\sim 7 \min$ \\
        The fitting formula of $\Lambda_t$ in Ref.~\cite{RN195} &  1.49 & perform direct calculations\\
        Neural network model &  1.49 &  $< 0.1$ s\\ \botrule
\end{tabular}}
\end{table}

\begin{figure}
    \centering
    \begin{overpic}
        [width=0.68\textwidth]{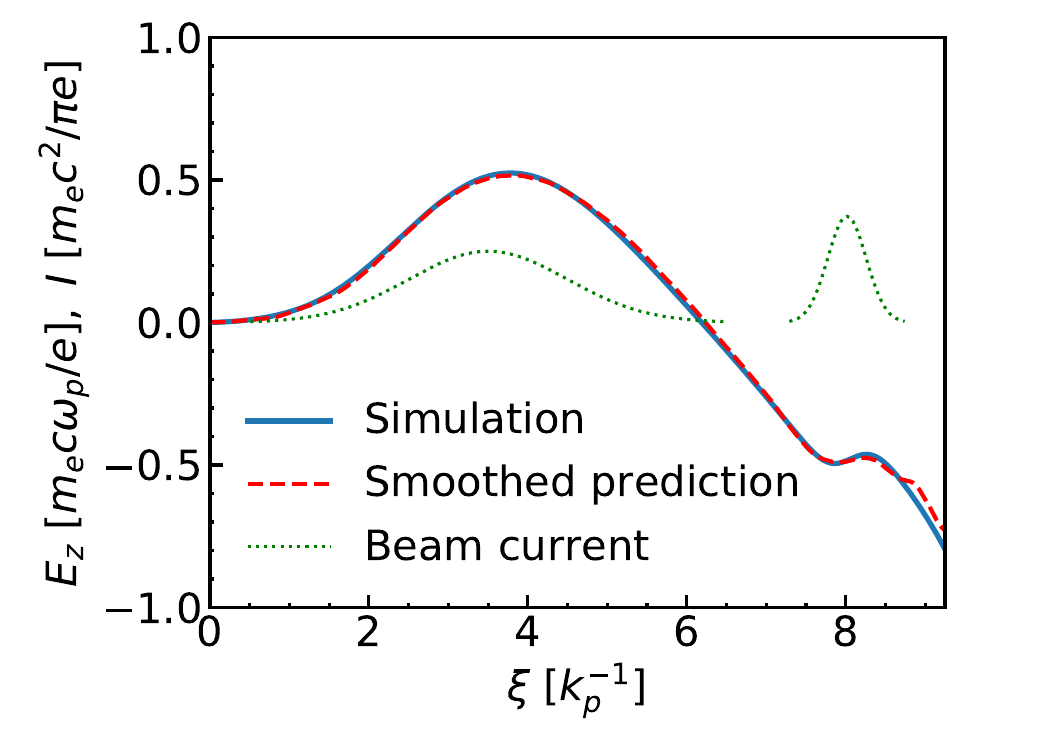}
    \end{overpic}
 \caption{Schematic of the beam current profiles and the (predicted) longitudinal wakefield distribution. The blue solid line represents the wakefield distribution curve from the PIC simulation, and the red dashed line is the smooth predicted line. The green dashed lines on the left and right represent the current distribution of the drive beam and the trailing beam, respectively.}\label{lam_NN_pre}
\end{figure}

In this section, we present another method that can enhance the optimization efficiency and provide a more visualized approach when designing a two-bunch PWFA. We utilize the trained neural network model to calculate the target $F(\Lambda_t)$ instead of QuickPIC during the optimization process, and to quickly obtain the optimal $\Lambda_t$ for the optimal beam-loading. Moreover, the corresponding longitudinal wakefield distribution can be directly obtained without performing PIC simulations. During the optimization, for the above case with input parameters $\Lambda_d=1.0$, $\sigma_{zd}=1.0$, $\sigma_{zt}=0.25$, $d=4.5$, $C_d=3.5$ and $L_z=9.25$, the optimal $\Lambda_t=1.49$ can be obtained within 0.1 seconds with the application of the neural network model. This case is not included in the training and validation datasets used in the model training. As shown in Tab.~\ref{lam_diff}, this result is consistent with the optimization using QuickPIC and the calculation using the fitting formula of $\Lambda_t$ in Ref.~\cite{RN195}. As shown in Fig.~\ref{lam_NN_pre}, the smoothed prediction by the neural network model is in good agreement with the simulated one. In the previous work~\cite{RN195}, the average time to obtain the optimal $\Lambda_t$ based on the PIC simulation is approximately 7.6 minutes. When using the neural network model to achieve the optimal beam-loading, the time consumption for different cases is within 0.1 seconds. This is because there is no need to partition the simulation grid as is required in PIC simulations, and the input-output dimensions are the same. This result indicates that the model can considerably reduce the optimization time and can effectively improve the design efficiency for PWFAs. The time comparison between the neural network model and PIC simulations is conducted by varying only one independent variable (the approach to obtaining $E_z$) while other conditions remain the same. 

The transformer ratio $R$ is also a crucial parameter that we care about in a PWFA design. In the following, we introduce a method based on the neural network model for quickly obtaining $R$, which is more intuitive than the fitting formula of $R$ proposed in Ref.~\cite{RN195}. The above parameters of achieving the optimal beam-loading $\Lambda_d=1.0$, $\sigma_{zd}=1.0$, $\sigma_{zt}=0.25$, $d=4.5$, $\Lambda_t=1.49$, $C_d=3.5$ and $L_z=9.25$ are taken as input variables of the model. Drive and trailing beams have a tri-Gaussian profile with the beam current distribution shown as green dashed lines in Fig.~\ref{lam_NN_pre}. Through the neural network model we trained, we can directly obtain the on-axis longitudinal wakefield distribution corresponding to the input variables given above. Furthermore, by combining the smoothed wakefield distribution with the beam current distribution, we can directly obtain the specific values of the maximum decelerating wakefield $W_{\rm dec}$ within the drive beam and the average accelerating wakefield $W_{\rm acc} = \int_{\xi_s}^{\xi_e} E_z(\xi)\lambda_{bt}(\xi)d\xi/\int_{\xi_s}^{\xi_e} \lambda_{bt}(\xi)d\xi$ within the trailing beam. Finally, by using the definition $R = \lvert W_{\rm acc}/W_{\rm dec}\rvert$, we can quickly obtain the transformer ratio $R$. Tab.~\ref{R_diff} shows the transformer ratio obtained through PIC simulations, the fitting formula of $R$ in Ref.~\cite{RN195}, and the neural network model we trained. The results obtained by the three methods are consistent, with no more than 2\% differences. Compared with the fitting formula of $R$ in Ref.~\cite{RN195}, the neural network model can provide more wakefield information. Thus, the effectiveness of the neural network model is verified. 

\begin{table}\small
\caption{\label{R_diff}
Transformer ratio $R$ and wakefield information obtained from the PIC simulation, the fitting formula of $R$ in Ref.~\cite{RN195} and the neural network model.}
\renewcommand{\arraystretch}{1.3}
{\begin{tabular}{C{0.40\textwidth}C{0.18\textwidth}C{0.18\textwidth}C{0.18\textwidth}} \toprule
        Method & $W_{{\rm dec}}$ & $W_{{\rm acc}}$ & $R$\\ \colrule
        PIC simulation & 0.524 & 0.495 & 0.945\\
        The fitting formula of $R$ in Ref.~\cite{RN195}& —— & —— & 0.948\\
        Neural network model & 0.520 & 0.482 & 0.927\\ \botrule
\end{tabular}}
\end{table}

\section{Conclusion}
We present a machine learning method based on the neural network to obtain the on-axis longitudinal wakefield distribution without the need for PIC simulations for the two-bunch PWFAs. We further explore the application of this model for optimizing the PWFAs. A method for achieving optimal beam-loading was proposed in a previous study~\cite{RN195}. By using BFGS algorithm in combination with the neural network model instead of the PIC simulation for feedback calculation, the optimization time can be reduced from approximately 7.6 minutes to less than 0.1 seconds. Although the fitting formula of $\Lambda_t$ in Ref.~\cite{RN195} can also be used for quickly obtaining the optimal $\Lambda_t$ without performing PIC simulations, the method proposed in this paper can directly provide the corresponding longitudinal wakefield distribution and improve the efficiency of PWFA design and optimization. Moreover, by utilizing the model given in this paper, additional information on the maximum decelerating wakefield within the drive beam and the accelerating wakefield within the trailing beam can be directly obtained. Subsequently, the transformer ratio can be determined. 

The optimization effectiveness of this model, especially when using diverse initial parameters, may be affected by variations in optimization cutoff accuracy and step size. This issue could be a potential avenue for future researches. The parameter space of this model, which is listed in Tab.~\ref{NN_tab}, covers most of the present PWFA facilities. Since the dataset used for model training is derived from the data during the process of achieving the optimal beam-loading, the model is primarily suitable for giving the wakefield information in the vicinity of the optimal beam-loading conditions.

\begin{acknowledgments}
This work is supported by the Strategic Priority Research Program of the Chinese Academy of Sciences (Grant No.~XDB0530000), and the National Natural Science Foundation of China (Grant No.~12475159 and No.~12075030).
\end{acknowledgments}

\bibliography{main}

\end{document}